\documentclass[useAMS,usenatbib]{mn2e}
\usepackage{graphicx}

\title[Precision Astrometry of Speckle Binaries with AO]
{Precision Astrometry of a Sample of Speckle Binaries and Multiples 
with the Adaptive Optics Facilities at the Hale and Keck II Telescopes}
\author[K. G. He{\l}miniak, M. Konacki, S. R. Kulkarni and J. Eisner]
{K. G. He{\l}miniak$^{1}$\thanks{E-mail:xysiek@ncac.torun.pl}, 
M. Konacki$^{1,2}$, S. R. Kulkarni$^{3}$ and J. Eisner$^{4}$
\\
$^{1}$Nicolaus Copernicus Astronomical Center, Department of Astrophysics, ul. Rabia\'{n}ska 8 , 
87-100 Toru\'{n}, Poland\\
$^{2}$Astronomical Observatory, A. Mickiewicz University, Sloneczna 36, 
60-286 Poznan, Poland\\
$^{3}$Division of Physics, Mathematics and Astronomy, California Institute of Technology, 
Pasadena, CA 91125, USA\\
$^{4}$Steward Observatory, University of Arizona, Tucson, AZ 85721, USA\\
}

\begin{document}

\date{Accepted  . Received  ; in original form  }

\pagerange{\pageref{firstpage}--\pageref{lastpage}} \pubyear{2008}

\maketitle

\label{firstpage}

\begin{abstract}
Using the adaptive optics facilities at the 200-in Hale and 10-m Keck II,
we observed in the near infrared a sample of 12 binary and multiple stars 
and one open cluster. We used the near diffraction limited images of these 
systems to measure the relative separations and position angles between 
their components. In this paper, we investigate and correct for the influence 
of the differential chromatic refraction and chip distortions on our relative 
astrometric measurements. Over one night, we achieve an astrometric  
precision typically well below 1 miliarcsecond and occasionally as small 
as 40 microarcseconds. Such a precision is in principle sufficient to astrometrically 
detect planetary mass objects around the components of nearby binary and multiple 
stars. Since we have not had sufficiently large data sets for the observed 
sample of stars to detect planets, we provide the limits to planetary mass 
objects based on the obtained astrometric precision.
\end{abstract}


\section{Introduction}

In the field of exoplanets, astrometry has not had too many triumphs to 
date. Of over 350 known planets or planetary candidates only one has been 
discovered astrometrically \citep{pra09}. The astrometric results presented by 
\citet{han01} are disputable and the true masses of only a few planets 
were calculated more reliably by combining the radial velocities (RV) and 
astrometry from the {\it Hubble Space Telescope} and the ground-based 
observations \citep{ben02, ben06}. Nevertheless, astrometry may just turn
out to be the most promising planet detection method in the future.
Astrometric space missions, like SIM \citep{unw08} or {\it Gaia} \citep{per05},
and a few ground-based interferometric surveys are ongoing \citep[e.g.][]{lane04} 
or in preparation \citep[e.g. on the VLTI:][]{eis08,lau08,sah08}. In particular 
ground based interferometers seem to be already well suited to detect planets 
by providing microarcsecond ($\mu$as) astrometric precision \citep[e.g.][]{mut05} for 
bright nearby binary stars.

The milliarcsecond (mas) or better precision can be achieved by imaging with 
the adaptive optics (AO) systems. This was already demonstrated for two binaries, 
HD~19063 and HD~19994, observed with the VLT \citep{neuh06,roel08} and a globular 
cluster M5 observed with the Hale telescope \citep{cam09}. Such a precision can be 
reached by means of relative astrometry over a small field of view \citep{cam09}. 
To this end one needs to have at least one reference object not too far from
a science object. For this reason visual and speckle binaries become
a natural target for such measurements. Incidentally, the subject of the
existence of exoplanets in binary and multiple stars has become of 
significant interest \citep[e.g.][]{ragh06,egg07,mut07,mug09}. It is now
accepted that the detection or lack of planets in star systems will provide
additional constraints to our models of planet formation \citep[e.g.][]{hol99,nel00,lis04}.

In this paper, we present our observations of a sample of 12 binary and multiple 
stars and one open cluster obtained in 2002 with the AO facilities 
at the Hale and Keck II telescopes over the period of 7 months. We
investigate the influence of several systematic effects that have an impact
on the relative astrometry and demonstrate that by correcting for them one
can achieve a sub mas precision. Finally, we provide the limits to planetary mass 
objects around components of our target stars derived from the obtained astrometric 
precision.

\section{Observations}
\begin{table}
 \caption{Number of images of a particular object per night.\label{tab_num}}
 \scriptsize
 \begin{tabular}{lccccc}
 \hline
 Night/Tel. & 56 Per & GJ 195 & GJ 300 & GJ 352 & GJ 458 \\
 \hline
04 Mar./K &  58 & --- &  58 & --- &  --- \\
23 Apr./H & --- & --- & --- &  53 &  975 \\
23 Jun./H & --- & --- & --- & --- & 1060 \\
24 Jun./H & --- & --- & --- & --- &  685 \\
21 Aug./H & --- & 300 & --- & --- &  --- \\  
22 Aug./H & --- & 582 & --- & --- &  --- \\  
13 Nov./H & --- & 949 & --- & --- &  --- \\  
 \hline
 \hline
 Night/Tel. & GJ 507 & GJ 569 & GJ 661 & GJ 767 & GJ 860 \\
 \hline
04 Mar./K &  --- &  29 &  --- & --- &  --- \\
23 Apr./H &  949 & --- &  656 & --- &  --- \\
23 Jun./H & 1012 & --- &  454 & --- &  189 \\
24 Jun./H &  520 & --- &  800 & --- & 1166 \\
26 Jun./H &  --- & --- & 1250 & --- &  --- \\
21 Aug./H &  --- & --- &  750 & 569 &  600 \\  
22 Aug./H &  --- & --- &  636 & 746 &  507 \\  
13 Nov./H &  --- & --- &  --- & 745 &  584 \\  
 \hline
 \hline
 Night/Tel. & GJ 873 & GJ 873B & GJ 9071 & & NGC 6871 \\
\hline
23 Jun./H &  225 & 251 &  --- & &  510 \\ 
24 Jun./H &  510 & 497 &  --- & & 1010 \\ 
21 Aug./H &  200 & 200 &  750 & & 1083 \\ 
22 Aug./H &  200 & 200 &  513 & & 2131 \\ 
13 Nov./H &  300 & --- & 1246 & &  624 \\
 \hline
\end{tabular}
\smallskip\\
A faint, third component of the GJ 860 system was not always in the field of view
(due to the dithering). Separate columns for GJ 873 and GJ 873B are to distinguish 
between the images of the full triple system and the double secondary only. 'K' stands 
for Keck II and 'H' for the Hale telescope.
\end{table}

\subsection{Instrumentation}
The main instrument in our project was the 200-in Hale Telescope at the Palomar 
Observatory. We used PHARO ({\it the Palomar High Angular Resolution
Observer}, \citealt{hayw01}) camera with PALAO ({\it the PALomar Adaptive Optics}) 
system. PHARO uses a mosaic of four $512\times512$ HgCdTe HAWAII detectors for 
observations between 1 and 2.5 $\mu$m \citep{hayw01}. PALAO is an adaptive optics 
system mounted at the Cassegrain focus of the telescope. It employs a Shack-Hartman 
wavefront sensor and a Xinetics Inc. 349/241 active-element deformable mirror. 
PALAO's detailed description can be found on-line\footnote
{http://ao.jpl.nasa.gov/Palao/PalaoIndex.html}.

With Hale/PALAO we obtained about 30,000 images of our targets. The data were 
collected over 7 nights nights between April and November 2002. We used an imaging 
mode with 39.91 and mostly 25.10 mas/pix scale and K, K', Ks broad, as well as 
Br$\gamma$ and FeII narrow band filters. We also used a 1\%-transmission neutral 
density filter ND-1 for decreasing a flux from very bright stars. Dithering was 
carried out by shifting the observed position by $\sim 2$ as.

We also had one clear night at the 10-m Keck II telescope. Using its AO system
and NIRC2 ({\it the Near InfraRed Camera 2}) we obtained data for three 
targets and the total of about 600 images. NIRC2 is a mosaic of four $512\times512$ 
InSb Aladdin-3 detectors. For the observations we used 9.942 and 39.686 mas/pix 
scale and the J, K' and K-cont (narrow-band) filters. Dithering was
done using field rotation.

\begin{figure}
 \includegraphics[width=\columnwidth]{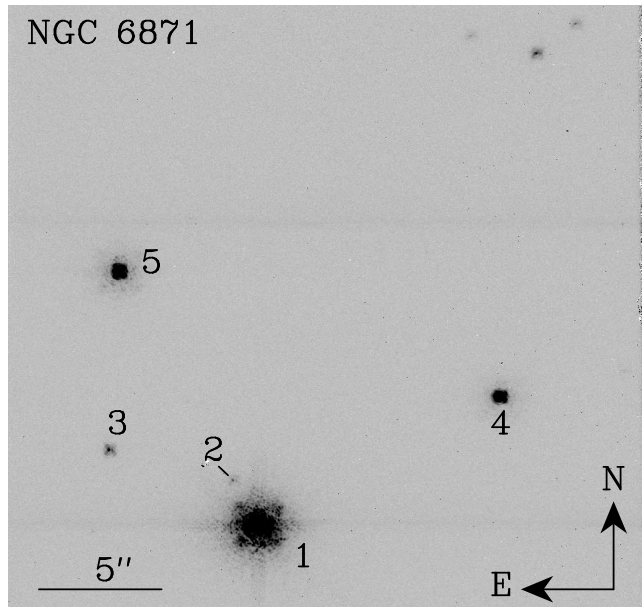}
 \caption{A field in NGC 6871 centreed on $\alpha=20^h 05^m 57^s, \delta=35^\circ 47' 25''$ 
as seen on 23 June 2002. Positions of only 5 labeled stars were measured. Other 3 (top of 
the image) are not seen in images from other nights. FOV is 25$\times$25 arcseconds. 
North is up, East is left. \label{img_ngc}}
\end{figure}

\begin{figure*}
 \includegraphics[width=0.87\textwidth]{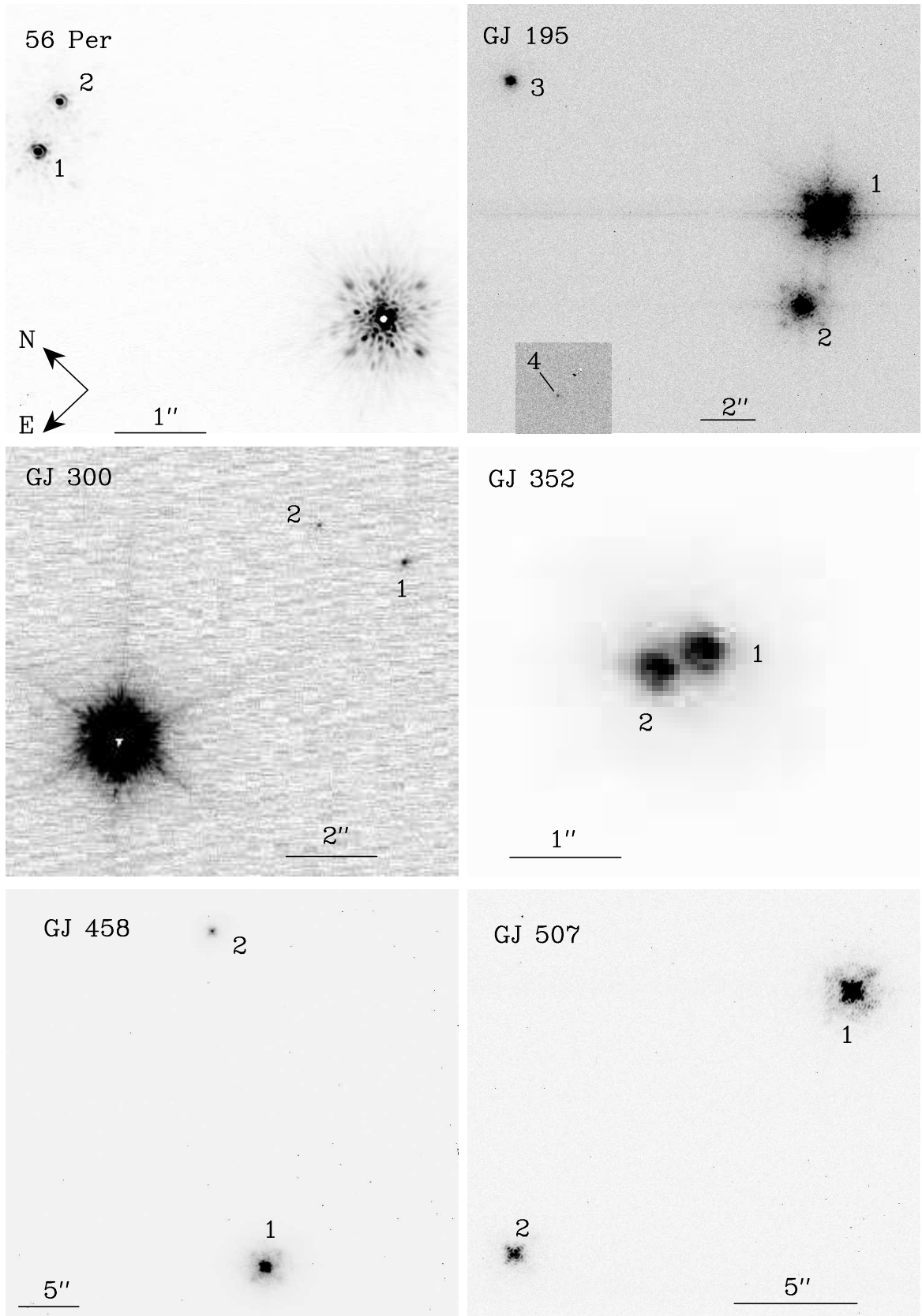}
 \caption{Images for all observed systems and labels of investigated stars. 
 North is up and East is left except for 56 Per where North is up-left 
 (rotation by 44.7 degrees, as shown). FOV size varies, separations are given in Table 
 \ref{tab_res}. The saturated cores of the primary components of 56 Per and GJ 300 are 
clearly  seen. GJ 569 A is also saturated but this is not seen in the adopted scale.
\label{fig_all}}
\end{figure*}
\begin{figure*}
 \begin{center}
 \includegraphics[width=0.87\textwidth]{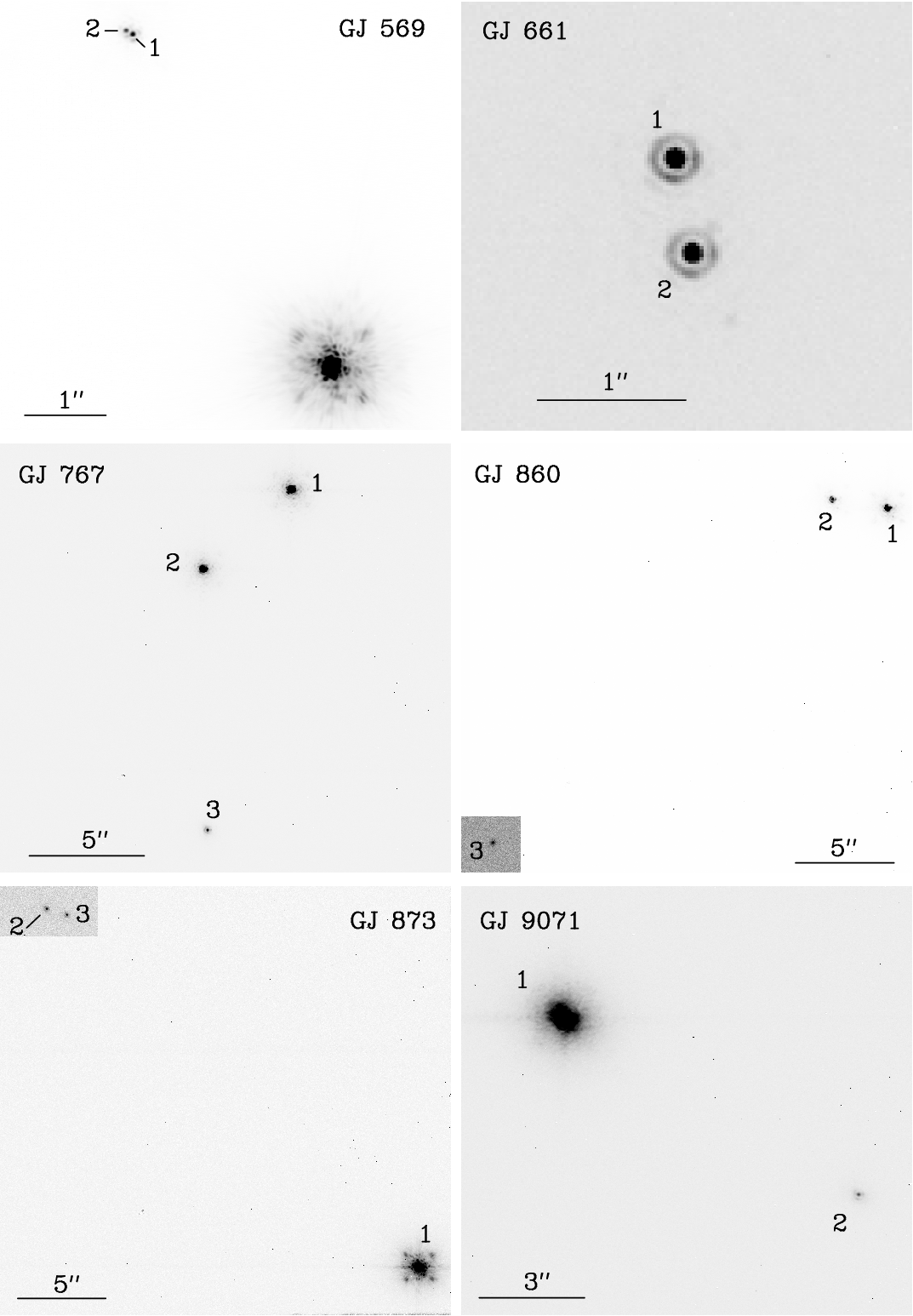}
 \contcaption{}
 \end{center}
\end{figure*}

\begin{table*}
 \caption{Basic information on our targets.\label{tab_prop}}
 \begin{tabular}{cccrlcccl}
  \hline
  Star & No. & Sp. Type &Magn.&(Band) & $\pi\,[mas]$ & Comment$^{a}$ & Telescope & Ref.\\
  \hline
  \hline
56 Per B  &1+2& ??? &8.7&(V)& 24.00(.91) &double&Keck II&1,2\\
GJ 195 A  &  1&M1   &10.16&(V)& 72.0(.4) & --- &Hale&3,4\\
GJ 195 B  &2  &M5   &13.7 &(V)& 72.0(.4) & --- &Hale&3,4\\
AG+45 517 &3  & ??? &11   &(V)& ??? & field &Hale&4\\
GJ 300 B  &1+2&K7III?&8.39 &(J)&125.60(.97)$^{b}$&double, field&Keck II&5,6\\
GJ 352 A  &1  &M4   &10.07&(V)& 94.95(4.31) & --- &Hale&1,7\\
GJ 352 B  &2  &M4   &10.08&(V)& 94.95(4.31) & --- &Hale&1,7\\
GJ 458 A  &1  &M0   &9.86 &(V) & 65.29(1.47) & --- &Hale&1,8\\
GJ 458 B  &2  &M3   &13.33&(V)& 65.29(1.47) & --- &Hale&1,8\\
GJ 507 A  &1  &M0.5 &9.52 &(V) & 75.96(3.31) & --- &Hale&1\\
GJ 507 B  &2  &M3   &12.09&(V)& 75.96(3.31) & --- &Hale&1\\
GJ 569 Ba &1  &M8.5V&11.14&(J)& 101.91(1.67) &double(?)$^{c}$&Keck II&1,9,10,11\\
GJ 569 Bb &2  &M9V  &11.65&(J)& 101.91(1.67) & --- &Keck II&1,9,11\\
GJ 661 A  &1  &M3   &10.0&(V) & 158.17(3.26) & --- &Hale&1,7\\
GJ 661 B  &2  &M4   &10.3&(V) & 158.17(3.26) & --- &Hale&1,7\\
GJ 767 A  &1  &M1   &10.28&(V)& 74.90(2.93) & --- &Hale&1,8\\
GJ 767 B  &2  &M2   &11.10&(V)& 74.90(2.93) & --- &Hale&1,8\\
GJ 860 A  &1  &M3   &9.59&(V) & 249.53(3.03) &variable&Hale&1,12\\
GJ 860 B  &2  &M4   &10.30&(V)& 249.53(3.03) &flare&Hale&1,12\\
CCDM 22281...H$^{d}$&3&???&13.8&(V)& ??? & field &Hale&13\\
GJ 873 A  &1  &M3.5e&10.09&(V)& 198.07(2.05) &flare&Hale&1\\
GJ 873 B  &2+3&G&10.66&(V)& 198.07(2.05) &double, field&Hale&1,14\\
GJ 9071 A &1  &K7&10.2&(V)& 72(4) & --- &Hale&1,8,13\\
GJ 9071 B &2  &M0&14&(B)& 72(4) & --- &Hale&1,13\\
  \hline
 \end{tabular}
\smallskip
\flushleft 
$^{a}$ If 'double', magnitude refers to a total magnitude of both
components and spectral type is 'averaged'. If 'field', the 
star is not gravitationally tied with brighter components.\\
$^{b}$ Parallax is for GJ 300, not the two investigated stars.\\
$^{c}$ \citet{sim06} suggested that GJ 569 Ba may be a binary with
similar brightness components.\\
$^{d}$ CCDM 22281...H = CCDM J22281+5741H -- a part of a multi-stellar 
system which includes also GJ 860.\\
Ref.:(1) The {\it Hipparcos} Catalogue \citep{per97}; (2) \citealt{bar05}; 
(3) \citealt{jen52}; (4) The {\it PPM North} Catalogue \citep{roe88}; (5) \citealt{sim96}; 
(6) \citealt{hen06}; (7) \citealt{alsh96}; (8) \citealt{rei04}; (9) \citealt{lane04}; 
(10) \citealt{sim06}; (11) \citealt{cut03}; (12) \citealt{law08}; 
(13) {\it CCDM - Catalog of Components of Double \& Multiple stars} \citep{dom02}; 
(14) \citealt{opp01} 
\end{table*}

\subsection{Objects}

From all the attempted objects we selected 9 binaries/multiples from the Hale 
data set and 3 multiples from the Keck II one. These are: GJ 195,
GJ 352, GJ 458, GJ 507, GJ 661, GJ 767, GJ 860, GJ 873 and GJ 9071 for the 
Hale, and GJ 300\footnote{Two fainter stars seen close to GJ 300 and GJ 873
are actually field stars.}, GJ 569 and 56 Per for the Keck II.  
All systems are shown in Figure \ref{fig_all}. We also selected 
a field in the open cluster NGC 6871 (the Hale data set) centreed 
around $\alpha=20^h 05^m 57^s, \delta=35^\circ 47' 25''$ as a 
reference field to study the systematic effects (Figure \ref{img_ngc}).

The selection criterion we adopted for the Hale sample was mainly the high 
number of individual images and also the number of nights during which a 
given object was observed. The exception is GJ 352. It was selected to study a 
precision of astrometry for a low number of single images. 
For the Keck II observations it was mostly important to check how good the 
AO correction was and how many unsaturated stars were in an image
(see Figure \ref{fig_all}). 
Due to saturation we were able to measure the relative positions only for the double 
secondary components of the Keck targets. The last two issues were
caused by the varying weather conditions and hence highly variable AO
correction. The final numbers of individual images per night  used in the
analysis for a particular object, after rejection of useless data, 
are given in Table \ref{tab_num}. 

Our objects are mostly M--type dwarfs located less than 20 $pc$ from the Sun. 
In a few cases (e.g. GJ 195) not only a binary but also other stars were 
captured in an image. Their relative positions were also measured. For 
the open cluster, the astrometry was performed only for 5 stars but other
objects can be seen in images as well. For all the stars which are cataloged 
the basic information is given in Table \ref{tab_prop}. Column 'No.' refers to 
a number/label in an image. The higher is the number assigned, the fainter 
the star is.

\section{Position calculations}

\begin{figure*}
 \includegraphics[width=0.75\textwidth]{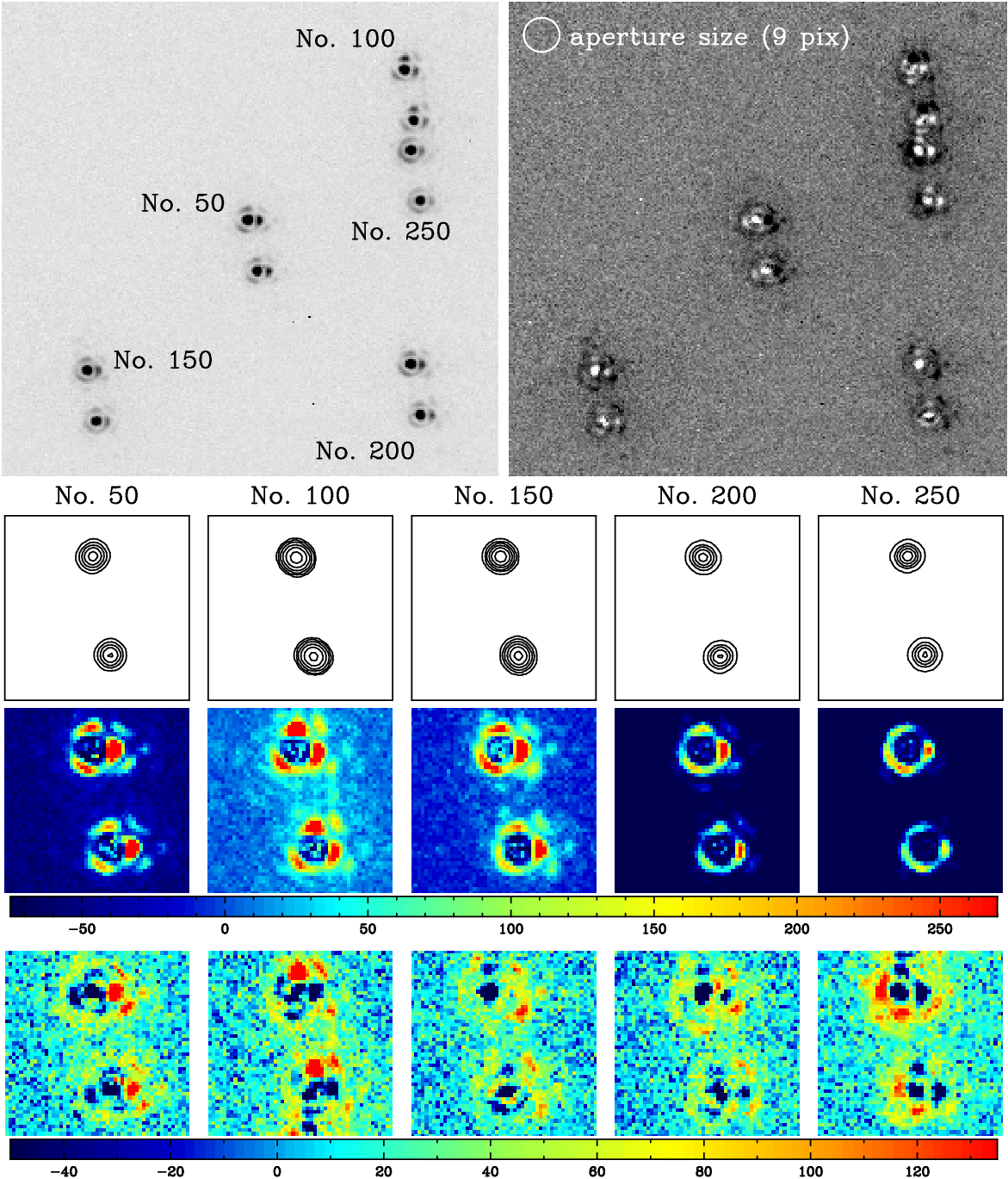}
 \caption{{\it Top left}: A combined image of GJ 661 from five exposures
 taken on June 23. The corresponding frame numbers are next to each 
 image of the binary;
 {\it Top right}: A residual image after a subtraction of an empirical PSF
 (computed with DAOPHOT; Stetson 1987).
 The size of the fitting aperture (9 pixels in radius) is shown as a white circle.
 {\it Upper middle:} A $15 \times 15$ pixel zooms on the contour plots of the fitted gaussian 
 functions at the positions of respective stars. Contour levels vary. Gaussians
 clearly show significant variations of their shape; {\it Lower middle:} The same  
 zooms on the residuals after a subtraction of the gaussian functions. Changes of the 
 first Airy ring can be seen. Color scale is the same in every sub-panel. 
 {\it Bottom:} The same zooms on the residuals after a subtraction of an empirical
 PSF. A clear leftover in the PSF core is evident. Color scale is the same in every 
 sub-panel. 
 \label{fig_psf}}
\end{figure*}

\begin{table*}
\caption{Parameters of an elliptical gaussian function used to model the images 
of stars in the frames for GJ 661 (Figure \ref{fig_psf}). An average value 
of a residual after a gaussian and an empirical PSF subtraction is given 
together with its uncertainties in the last four columns. Analyzed 
frames are numbered as in  Fig. \ref{fig_psf}.
\label{tab_par_gaus}}
\begin{tabular}{lcccccccccccc}
\hline 
Star & $A$ & $\sigma_x$ & $\sigma_y$ & $x_0$ & $y_0$ & $B$ & $\theta$ & & Av. resid. &
$\sigma$ & Av. resid. & $\sigma$ \\
Frame/ID & [counts] & [pix] & [pix] & [pix] & [pix] & [counts] & [$^\circ$] & & (gauss) & & (PSF) & \\
\hline \hline
No. 50/1  & 2846.443 & 1.505 & 1.432 & 543.897 & 615.032 &  13.231 & 130 & & 17.59 & 18.16 & -111.73 & 13.57 \\
No. 50/2  & 2203.045 & 1.437 & 1.429 & 548.635 & 588.264 &  48.128 &  91 & &  3.74 & 11.38 & -101.62 & 10.73 \\
No. 100/1 & 2711.196 & 1.399 & 1.499 & 622.879 & 694.661 &  -0.042 & 122 & &  4.88 & 11.46 &  -39.33 &  5.68 \\
No. 100/2 & 2125.978 & 1.378 & 1.512 & 627.583 & 667.851 &  -0.300 & 125 & & -3.95 & 14.32 &  -85.35 &  7.45 \\
No. 150/1 & 2629.679 & 1.478 & 1.407 & 626.141 & 538.949 &  35.679 &  16 & & 11.13 & 11.38 & -130.78 & 10.17 \\
No. 150/2 & 1953.843 & 1.636 & 1.527 & 630.955 & 512.050 & -17.370 &  65 & & 15.58 & 17.12 & -116.19 & 17.75 \\
No. 200/1 & 2391.502 & 1.395 & 1.488 & 462.866 & 535.715 &  59.381 & 115 & &  3.06 & 10.35 & -140.15 &  8.43 \\
No. 200/2 & 1928.833 & 1.363 & 1.469 & 467.544 & 508.818 &  51.934 &  65 & &  8.88 & 13.26 &  -75.02 &  7.67 \\
No. 250/1 & 2344.759 & 1.436 & 1.349 & 626.177 & 652.063 &  81.791 & 165 & &  2.47 &  7.87 & -112.96 &  9.03 \\
No. 250/2 & 1722.767 & 1.441 & 1.450 & 630.861 & 625.335 &  50.023 &  87 & &  0.83 &  6.09 &  -67.74 &  7.36 \\
\hline
\end{tabular}
\end{table*}

The images were first reduced with standard IRAF tasks for data reduction. 
Subsequently, given the number of exposures, the relative positions of 
the stars were computed with our own software in an automated 
manner as follows: (1) the shifts from 
image to image (due to the dithering) were measured by cross-correlating the 
template image (usually the first image) with all the subsequent exposures and 
the approximate positions of stars in a given image were calculated with an 
accuracy of $\pm 3$ pix ($\sim 75$ mas for most of the Hale's sample), 
(2) based upon these positions, the centroids were calculated, (3) based upon 
the centroids, the following 2-dimensional elliptical gaussian function was 
fitted to the cores of the images of stars:
\begin{eqnarray}
\begin{array}{rr}
&G(x,y) = B + A \exp{}\left[ -\frac{[(x-x_0)\cos{\theta} - (y-y_0)\sin{\theta}]^2}{2\sigma^2_x} \right. \\
&\left. - \frac{[(x-x_0)\sin{\theta} + (y-y_0)\cos{\theta}]^2}{2\sigma^2_y} \right].\\
\end{array}
\end{eqnarray}
where $B$ is a background level, $A$ the amplitude of a gaussian, $(x_0,y_0)$
is a position of the star, $\sigma_x,\sigma_y$ the correspondimg widths
and $\theta$ is a tilt of the gaussian.

We have decided to use such an approach because it offers a simple and robust 
way of modeling of the cores of stars' images. One could envision using an empirical
point spread function (PSF) as a model for the images of stars. This is however 
challenging due to a 
fact that in a single image we typically have only two stars. Hence our
knowlegde about the actual empirical PSF for a given frame is limited.
Additionally, since the PSF's shape varies, it is not practical to use
several subsequent exposures as a reference for an averaged empirical.
This is demonstrated in Figure \ref{fig_psf} for a series of 5 frames of GJ~661 
taken on June 23 and spanning 15.5 minutes for which an average empirical PSF 
is calculated using a 9 pixel aperture and subtracted from the images of stars.
As can be seen a fitting of the gaussian performs better. The details
of this procedure are also given in Table \ref{tab_par_gaus}.

The results of the gaussian fitting were 
used to compute the relative separations and the position angles of pairs of 
stars. The NIRC2 data were corrected for the field rotation used for
dithering. Let us note that we did not use any weighting scheme for individual
images as it was done by \citet{cam09}. A significant improvement in the astrometric 
precision after using an optimal weighting is seen when the number of reference
stars exceeds 5 \citep[see Fig. 2 in][]{cam09} which is never the case in 
our work targeting binary stars and aimed at investigating the astrometric
precision in the case of only one reference star.

\section{Systematic effects}\label{sec_sysef}

\subsection{Adaptive Optics correction and field of view}

\begin{figure*}
 \includegraphics[angle=270.,width=0.88\textwidth]{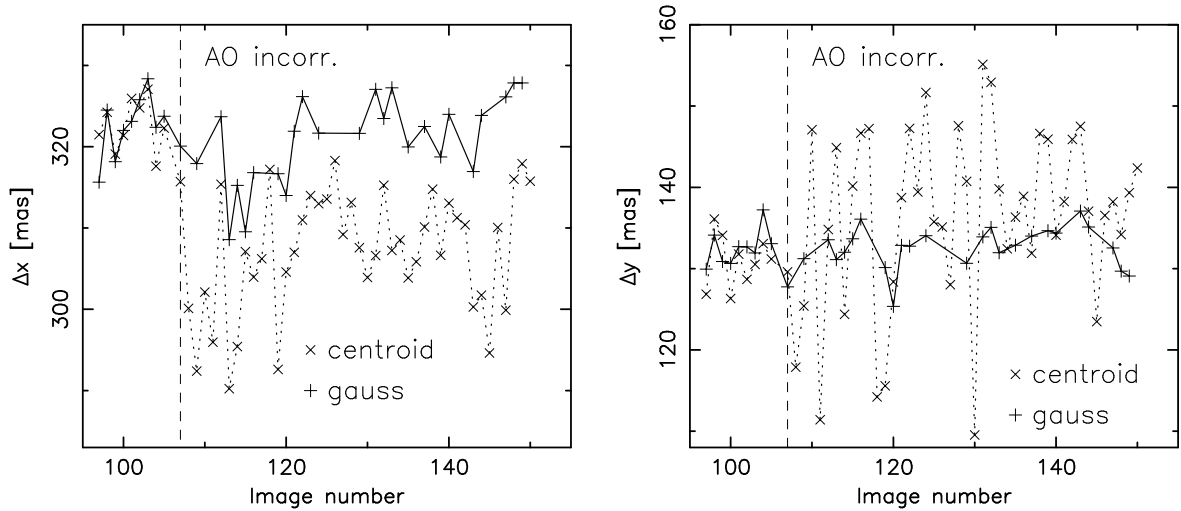}
 \caption{Impact of Adaptive Optics correction on relative position 
measurements in case of GJ 352. After image No. 107 (dashed line) AO 
works improperly.\label{352_AO}}
\end{figure*}


The main factor allowing us to obtain precise astrometric measurements 
is obviously the adaptive optics. Its performance will influence the
final astrometric precision as can be seen in the case of GJ 352 ($\rho
\simeq 350$ mas) observed during challenging weather conditions. 
From all 75 images of GJ 352 taken only the first 10 were properly 
corrected (the Airy pattern visible) and only in 53 images components 
were resolved and could be analysed. For these 53 images 
the centroids were calculated. Subsequent gaussian fitting was possible 
only for 34 images for which the fitting procedure converged. 
The centroids and the outcome of gaussian fitting 
are in agreement for the first 10 (Figure \ref{352_AO}).

Another factor having an impact on the astrometric precision is the 
field of view and the corresponding PSF sampling which was especially 
important for the Keck II data. The images were taken in two 
pixel scales (field sizes) -- 9.942 mas/pix ($10''\times10''$ field, narrow) 
and 39.686 mas/pix ($40''\times40''$ field, wide) and in various field rotator 
angles -- $45.7$, $0.7$ and $-44.3$ deg. For this telescope the 
diffraction-limited size of a star's image in 
the near infrared corresponds to about 1.4 pixel in the wide field. For faint 
stars it means that most of their light is collected in one pixel which makes 
the PSF undersampled and the gaussian fitting difficult. This issue, however, 
can be at least partially overcame by a sub-pixel dithering. For the purpose of 
this paper, we used only the frames taken with the 'narrow' field.

\subsection{Atmospheric refraction}

The atmospheric differential refraction (ADR) creates a shift of star's 
image. It is highly dependent on the zenith angle and wavelength. Formulae for 
computing ADR effect are given for example by Roe (2002) where the angle $R$,
which is the difference between the real and observed zenithal distance is
given by:

\begin{equation}\label{eq_R}
R\equiv z_t - z_a \simeq 206265\left( \frac{n^2-1}{2n^2}\right)\tan{z_t} \ [arcsec]
\end{equation}
where $z_t$ is the true zenithal distance, $z_a$ is the observed zenithal distance, 
and $n$ is the refraction index, dependent on the wavelength $\lambda$ and weather 
conditions:

\begin{eqnarray}\label{eq_n}
\begin{array}{l}
n(\lambda,p,T,p_w)=1\\
\\
+\left[64.328+\frac{29498.1}{146-\lambda^{-2}}+\frac{255.4}{41-\lambda^{-2}}\right]\frac{pT_s}{p_sT}10^{-6}\\
\\
-43.49\left[1 - \frac{0.007956}{\lambda^2}\right]\frac{p_w}{p_s}10^{-6}
\end{array}
\end{eqnarray}
where $\lambda$ is given in $\mu$m, \textit{p, T} and $p_w$ are the pressure 
[hPa], temperature [K] and partial pressure of water vapor respectively. 
Symbols with the index $s$ refer to the canonical values of air pressure (1013.25 
hPa) and temperature (288.15 K). The angle $R$ is much smaller in IR than 
in visible. 

ADR also affects relative astrometric measurements. Since two objects are seen
at different zenithal distances $z_1$ and $z_2$, the corrections $R_1$ and $R_2$ 
are also different. The component of the separation vector parallel to the 
direction to the zenith increases after ADR correction by $\Delta R =|R_2 - R_1|$.
This quantity changes with weather conditions (air pressure and temperature). As 
we have demonstrated \citep{hel09}, the magnitude of this change is 
often higher than an achievable astrometric precision even for relatively 
compact systems. Clearly, ADR's influence must be corrected for and the weather 
conditions should be well known. This conclusion is in contrary to that by 
\citet{neuh06}, who claim that the refraction is in general insignificant thanks to 
the use of a narrow bandpass filter (chromatic refraction). As the ADR is dependent 
on the zenithal distance, it may be significant even for a monochromatic light. 
So, the real reason why ADR is negligible in case of \citet{neuh06} is likely 
the geometry of their binary. Nevertheless, it is true that using wide-band filters 
makes ADR harder to calculate due to it's chromatic character and, for example, stars' 
different colors \citep{hel09}. 

Unfortunately, we did not collect any weather readings during our observing runs, 
so we had to use the canonical values of temperature and pressure and
assumed 50 per cent humidity. For the Keck observations we assumed 2 times smaller pressure 
and the temperature of $0^{\circ}$C. This means that the real uncertainties
of measured separations and position angles are higher than the precisions given 
in Table \ref{tab_res}. To correct for ADR, we used the semi-full approach
as described by \citet{hel09}.

In order to estimate the maximum error due to ADR, we took the largest possible 
separation in our sample -- 30.8 as for GJ 873 1-2 (from Table \ref{tab_res}). 
We assumed that the 
difference in the zenithal angles equals to the separation and the binary is seen 30 degrees 
above the horizon. In such an improbable case the maximum contribution to the error budget 
coming from the temperature is 4 mas (if the real temperature
were $T$ = 230 K and pressure $p$ = 1013.25 hPa) and the fraction coming from the air pressure is smaller 
than 8 mas (for $p$ = 613 hPa and $T$ = 230 K). One should add that the bigger is the 
part coming from the temperature, the smaller is the contribution from the pressure. So we 
may conclude that in this improbable case 8 mas is the maximum error and in most (if not 
all) our real observations the uncertainty caused by the weather conditions is smaller than 
several mas. In the case of binaries observed with Keck II, the maximum error should be 
much smaller because the separations (and $z$ differences) are smaller and other, more
probable weather conditions were assumed. The maximum uncertainty scales linearly with
air pressure and almost linearly with separation (maximum $z$ difference) and temperature.
In Section 5.1 we estimate in yet another way the observed rms of our
astrometric measurements.

\subsection{Chip geometry and orientation}

\begin{table}
\caption{Average pixel scales and North orientations for Hale telescope.\label{tab_psc}}
\begin{tabular}{lcc}
\hline
Night & Scale & North \\
      & (mas/pix) & (degrees) \\
\hline
\hline
23 Apr. (narrow)& 25.168(34) & 334.043(99) \\
23 Apr. (wide)&  40.00(2)  & 334.043(99) \\
23 Jun. & 25.168(34) & 334.043(99) \\
24 Jun. & 25.171(21) & 334.039(18) \\
26 Jun. & 25.171(21) & 334.039(18) \\
21 Aug. & 25.156(10) & 334.072(11) \\
22 Aug. & 25.156(10) & 334.072(11) \\
13 Nov. & 25.156(10) & 334.723(15) \\
\hline
\end{tabular}
\end{table}

\begin{figure*}
\includegraphics[width=0.91\textwidth]{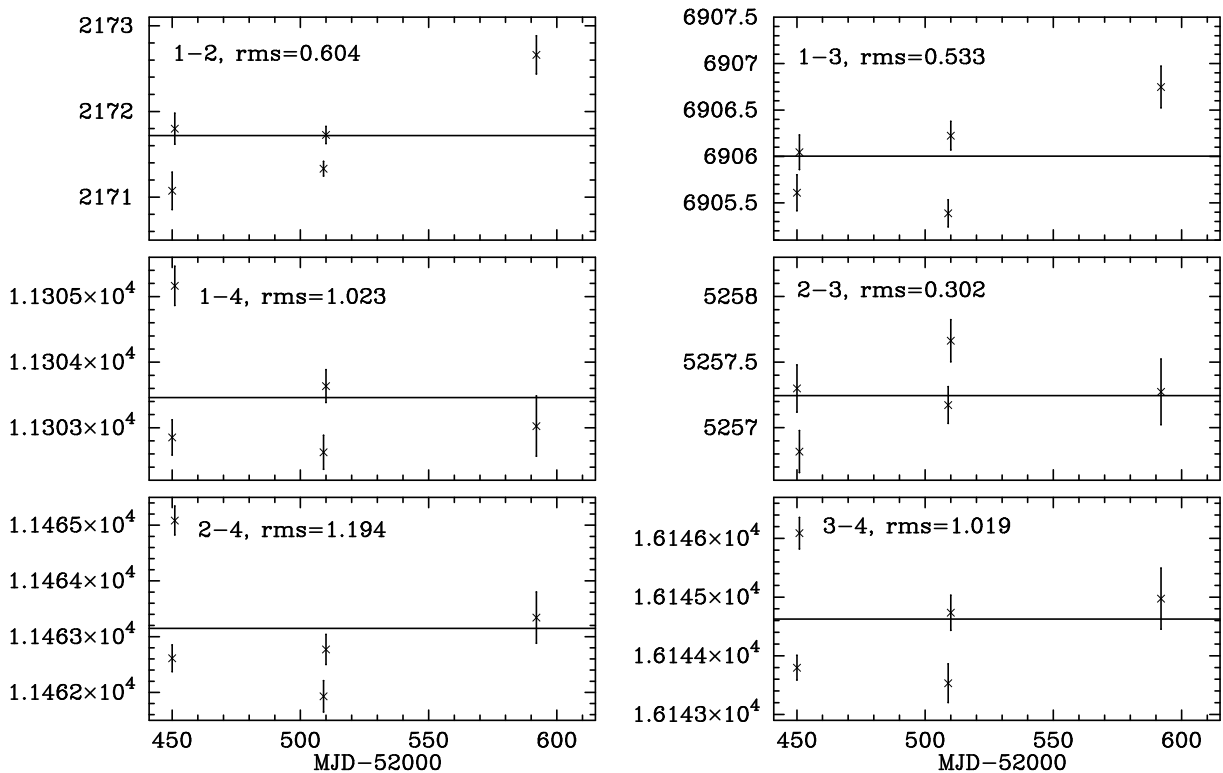}
\caption{Separations of stars from the open cluster NGC 6871. The average value is
plotted as a solid line and the corresponding $rms$ in mas is shown.\label{fig_sepfit}} 
\end{figure*}

\begin{table}
\caption{The average separations between the stars 1 to 4 in the NGC 6871 open cluster.
\label{tab_sepfit}}
\begin{tabular}{rrl}
\hline
Pair & $\rho$ [mas] & $rms$ \\
\hline \hline
NGC 6871 1-2 & 2171.719 & 0.604 \\
	1-3 &  6906.003 & 0.533 \\
	1-4 & 11303.461 & 1.023 \\
	2-3 &  5257.245 & 0.302 \\
	2-4 & 11463.147 & 1.194 \\
	3-4 & 16144.625 & 1.019 \\
\hline
 \end{tabular}
\end{table}

Detectors are not perfectly rectangular, flat and perpendicular to the 
light-path. At the astrometric precision necessary to detect planets (well below 
1 mas), one has to know how the camera's detector is distorted or how the
pixel scale and the detector's orientation changes from epoch to epoch. For 
instruments mounted in the Cassegrain focus, as is the case for PHARO, the 
distortion changes with the telescope's position due to gravity. This effect 
is not present or negligible in the case of NIRC2 which is located in a Nasmyth 
platform. The astrometric calibration and distortion models are available for both 
cameras. The calibration for PHARO is more complicated and includes not only 
the geometry and orientation of the detector itself but also the influence of 
telescope's position and tilt of the chip relatively to the light path. 
This is described by \citet{met06}\footnote{See http://www.astro.ucla.edu/$\sim$metchev/ao.html}. 
The distortion of the NIRC2 camera was investigated during it's pre-ship testing 
and is described by \citet{thom01}\footnote{See http://alamoana.keck.hawaii.edu/inst/nirc2/preship\_testing.pdf}.

The calibration we carried out included deriving the average plate-scale of the chip 
and the position of the North direction with respect to the Cassegrain ring (CR). 
The nominal values are 25.10 mas/pix for the narrow, 39.91 mas/pix for the wide 
field pixel scale and 335$\fdg$8 for CR \citep{hayw01}. As it was shown by 
\citet{met04}, the real values are different from the nominal one and usually change 
from epoch to epoch. As a base for our calibration, we adopted the measurements
from 23 June 2002 by \citet{met04} which are: $25.168\pm 0.034$ mas/pix and 
334.043$\pm 0\fdg099$. We chose 4 stars in NGC 6871, marked in Fig. \ref{img_ngc} 
as 1-4 which we believe to be members of the cluster (their positions with
respect to the star No. 5 changed in a similar way), and using their 
relative positions we have recalculated the average pixel scale in the narrow field
by assuming that their astrometric motion is not detectable.
First of all, for every night and every of six possible pairs, we calculated a preliminary
pixel scale and the North direction incorporating the uncertainties from \citet{met04}.
Later we averaged the results for every single night. We checked if for two or more 
consecutive nights the pixel scale changed, and if it did not, we averaged the result 
over the number of nights. This procedure allowed us to improve our pixel
scale's uncertainties with respect to those given by \citet{met04} who used only one pair
while we used up to 18 (six pairs, observed during 3 nights).
An adequate procedure was carried out for the CR orientation angle. 

For 3 nights of August and November 2002 we obtained the value of $25.156 \pm 0.010$ 
mas/pix which is in poor agreement with the previous value but then the difference in 
separations of the stars was clearly seen. We also recalculated new pixel scale for June 24 
and obtained 25.171 $\pm$ 0.021 mas/pix which shows that there was actually no scale 
change during one night. We also assumed the same plate-scale for the night of June 26. 
The cluster NGC 6871 was not observed in April 2002 but 
the results for GJ 458 suggest that the pixel scale and orientation was the same 
as in June. We assumed that the orbital period of GJ 458 is long enough
that no orbital motion can be seen after 62 days and the separation remains constant.
Comparing this system with GJ 195 which is closer to the Sun and has a few times smaller 
angular separation, we may conclude that the period is much longer than 338 years 
(orbital period of GJ~195, \citealt{hein74}), probably close to 3200 years. 
We see only a small (however clear) motion in GJ 195 and we should expect at least 
a 10 times smaller movement in GJ 458 which would be below the detection. Thus for 
April 2002 we used the same pixel scale as for June 23 with its relatively 
high uncertainty.
 
This binary was also observed in the wide field mode on 23 April at the beginning 
of the night. By combining the measurements from the wide and narrow fields, we 
obtained $40.00 \pm 0.02$ mas/pix as a pixel scale for the wide field and noticed 
no change in the CR angle. Between June and November we noticed two changes of the 
CR orientation. The values of CR for August and November were: $334.072 \pm 0.011$ 
and $334.723 \pm 0.015$ degrees respectively. The recalculated value for June 24 and 26
was $334.039 \pm 0.018$ degrees. We should also note that the position angles were 
computed from $\Delta x$ and $\Delta y$ counted in pixels along the chip's axes, so 
the bigger separation, the smaller uncertainty in $\theta$. 

Values of the derived plate-scale and North positions are summarized in Table \ref{tab_psc}. 
For the stars in the open cluster NGC 6871, we show the average values of
separations together with their $rms$' (Tab. \ref{tab_sepfit}, Fig. \ref{fig_sepfit}). 
The average values are calculated from 5 epochs corrected for the new pixel
scales. Such a calibration is not perfect and these plate-scale and North
direction values are
subject to possible systematic errors e.g. due to a limited knowledge of the 
weather conditions. 

\begin{figure*}
\includegraphics[width=\textwidth]{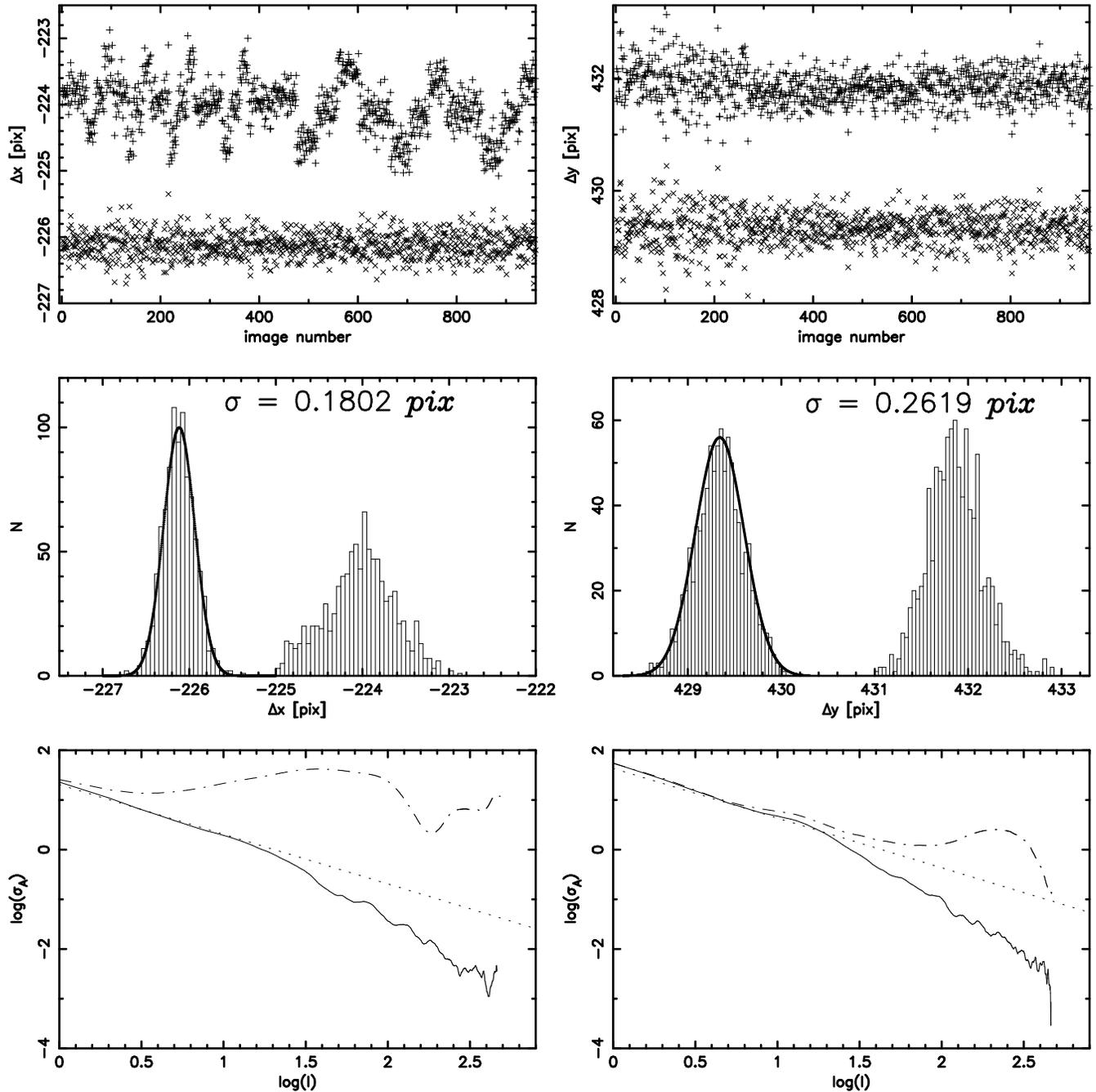}
\caption{An example of the distortion and its correction for two stars in NGC 6871 
cluster. Left panels refer to X ($\alpha$) component, right ones to Y ($\delta$). 
Top panels: the measurements before ($+$) and after ($\times$, shifted) 
the distortion correction. Middle panels: the histograms of the measurements 
(bin width of 0.05 $pix$) with a gaussian fitted to the corrected measurements
(left). Bottom panels: the Allan variance of the uncorrected (dot-dashed) and 
corrected (solid) measurements and an infinitely long, white-noise 
signal with $\sigma$ given in the middle panel (dotted line). \label{dist_corr}}
\end{figure*}

Examples of the distortion and the results of employing its model can be seen 
in Figure \ref{dist_corr}. We show the uncorrected and corrected measurements 
of separation ($\Delta x, \Delta y$) for two stars in the NGC 6871 field. 
The distortion is clearly seen in the $x$-axis (the top-left panel).
It is worth noting that for PHARO the distortion is much bigger in $x$ than in $y$, 
but in the $y$-axis the random scatter is about 50\% bigger. The histograms
demonstrate that after the correction, we are able to obtain a gaussian 
statistics (the middle panels) and the Allan variance (AV) shows no obvious 
signs of systematic errors (the bottom panels).
 
The real average plate-scales of NIRC2 were found to be in agreement with the nominal 
values \citep{met04} but the $y$-axis was rotated by 1$\fdg$24 clockwise from North 
\citep{met05}. Unfortunately, due to a small number of useful images in our
data set, we were not able to perform proper tests and our own calibrations. Hopefully, 
the location of the camera on the Nasmyth platform grants its stability.
In particular, the results for 56 Per and GJ 569 B where three different field rotator 
positions were used demonstrates that the precision of the field rotator of NIRC2 is 
better than $0\fdg1$ during one night.

\begin{table*}
\caption{Separations and position angles of investigated stars.
\label{tab_res}}
 \begin{tabular}{lrlrlrclrlrlr}
 \hline
Pair & $\rho \,[mas]$ & $\pm$ & $\theta \, [^\circ]$ & $\pm$ & MJD  & &
Pair & $\rho \,[mas]$ & $\pm$ & $\theta \, [^\circ]$ & $\pm$ & MJD \\
 \hline \hline
{\bf 56 Per B} &&&&&&&					{\bf GJ 767} &&&&&\\
1--2 & 626.31 & 0.32 & 291.733 & 0.039 & 52337&&	1--2 &  5276.64 & 0.17 & 135.4224 & 0.0014 & 52509 \\
 &&&&&&&						     &  5277.21 & 0.12 & 135.4129 & 0.0009 & 52510 \\
{\bf GJ 195} &&&&&&&					     &  5284.86 & 0.09 & 135.4630 & 0.0006 & 52592 \\
1--2 &  3612.79 & 0.31 & 167.3130 & 0.0012 & 52509 &&	1--3 & 16016.88 & 0.52 & 168.9361 & 0.0004 & 52509 \\
     &  3613.02 & 0.23 & 167.3020 & 0.0009 & 52510 &&	     & 16017.73 & 0.28 & 168.9188 & 0.0002 & 52510 \\
     &  3612.18 & 0.12 & 167.4167 & 0.0004 & 52592 &&	     & 15840.08 & 0.39 & 169.1386 & 0.0003 & 52592 \\
1--3 & 12830.54 & 0.50 &  67.7436 & 0.0017 & 52509 &&	2--3 & 11977.08 & 0.43 & 183.01454 & 0.00005 & 52509 \\
     & 12831.67 & 0.44 &  67.7327 & 0.0018 & 52510 &&	     & 11977.06 & 0.22 & 182.99594 & 0.00002 & 52510 \\
     & 12869.51 & 0.24 &  67.4695 & 0.0009 & 52592 &&	     & 11811.14 & 0.35 & 183.50265 & 0.00007 & 52592 \\
1--4 & 12151.22 & 0.54 & 126.6784 & 0.0020 & 52509 &&	 &&&&&\\
     & 12150.86 & 0.47 & 126.6756 & 0.0017 & 52510 &&	{\bf GJ 860} &&&&&\\
     & 12125.38 & 0.23 & 126.3902 & 0.0008 & 52592 &&	1--2 &  2875.849 & 0.078 & 84.0520 & 0.0014 & 52450 \\
2--3 & 13895.74 & 0.43 &  52.8881 & 0.0013 & 52509 &&	     &  2875.365 & 0.048 & 84.0420 & 0.0006 & 52451 \\
     & 13896.73 & 0.46 &  52.8783 & 0.0016 & 52510 &&	     &  2858.542 & 0.061 & 83.2373 & 0.0010 & 52509 \\
     & 13954.70 & 0.24 &  52.6982 & 0.0008 & 52592 &&	     &  2858.621 & 0.061 & 83.2063 & 0.0009 & 52510 \\
2--4 &  9699.14 & 0.51 & 112.6419 & 0.0024 & 52509 &&	     &  2834.493 & 0.087 & 82.0401 & 0.0014 & 52592 \\
     &  9698.21 & 0.47 & 112.6382 & 0.0022 & 52510 &&	1--3 & 26525.54 & 0.73 & 133.9775 & 0.0012 & 52450 \\
     &  9694.63 & 0.21 & 112.2336 & 0.0011 & 52592 &&	     & 26528.41 & 0.41 & 133.9754 & 0.0007 & 52451 \\
3--4 & 12303.86 & 0.50 & 189.9677 & 0.0034 & 52509 &&	     & 26748.78 & 0.94 & 133.5531 & 0.0016 & 52509 \\
     & 12305.32 & 0.60 & 189.9642 & 0.0004 & 52510 &&	     & 26755.53 & 1.37 & 133.5333 & 0.0025 & 52510 \\
     & 12310.07 & 0.24 & 189.9467 & 0.0002 & 52592 &&	     & 26851.07 & 1.34 & 132.4185 & 0.0019 & 52592 \\
 &&&&&&&						2--3 & 24771.78 & 0.75 & 139.0745 & 0.0012 & 52450 \\
{\bf GJ 300 B}&&&&&&&					     & 24775.50 & 0.42 & 139.0711 & 0.0007 & 52451 \\
1--2 & 2035.74  & 0.12 &  66.6688 &  0.0018 & 52337 &&	     & 25020.13 & 0.91 & 138.5973 & 0.0015 & 52509 \\
 &&&&&&&						     & 25028.25 & 1.19 & 138.5767 & 0.0021 & 52510 \\
{\bf GJ 352} &&&&&&&					     & 25139.60 & 1.33 & 137.4029 & 0.0019 & 52592 \\
1--2 &   346.21 & 1.11 & 113.6970 & 0.12 & 52389 &&	 &&&&&\\
 &&&&&&&						{\bf GJ 873} &&&&&\\
{\bf GJ 458} &&&&&&&					1--2 & 30155.85 & 0.57 &  47.0269 & 0.0009 & 52450 \\
1--2 & 14723.58 & 0.40 &  10.55272 & 0.00024 & 52389 &&      & 30158.79 & 1.07 &  47.0469 & 0.0016 & 52451 \\
     & 14720.19 & 0.28 &  10.56016 & 0.00016 & 52450 &&      & 30328.74 & 0.73 &  47.2967 & 0.0012 & 52509 \\
     & 14723.27 & 0.36 &  10.55398 & 0.00021 & 52451 &&      & 30334.65 & 0.60 &  47.3068 & 0.0009 & 52510 \\
 &&&&&&&						     & 30784.76 & 1.12 &  47.4235 & 0.0015 & 52592 \\
{\bf GJ 507} &&&&&&&					1--3 & 29089.44 & 0.58 &  45.8990 & 0.0009 & 52450 \\
1--2 & 17747.73 & 0.45 & 131.0180 & 0.0012 & 52389 &&	     & 29093.40 & 0.85 &  45.9219 & 0.0013 & 52451 \\
     & 17757.84 & 0.33 & 131.0927 & 0.0006 & 52450 &&	     & 29260.32 & 0.76 &  46.1888 & 0.0012 & 52509 \\
     & 17756.96 & 0.66 & 131.0845 & 0.0013 & 52451 &&	     & 29265.43 & 0.61 &  46.1990 & 0.0009 & 52510 \\
 &&&&&&&						     & 29717.16 & 1.03 &  46.3402 & 0.0014 & 52592 \\
{\bf GJ 569 B} &&&&&&&					2--3 &  1215.619 & 0.067 & 255.1128 & 0.0034 & 52450 \\
1--2 &    98.14 & 0.11 &  61.506 & 0.050 & 52337 &&	     &  1216.078 & 0.266 & 255.1578 & 0.0170 & 52451 \\
 &&&&&&&						     &  1214.913 & 0.085 & 255.0884 & 0.0044 & 52509 \\
{\bf GJ 661} &&&&&&&					     &  1215.301 & 0.100 & 255.0963 & 0.0048 & 52510 \\
1--2 &   724.611 & 0.079 & 195.3279 & 0.0015 & 52389 &&      &  1214.012 & 0.347 & 255.0787 & 0.0170 & 52592 \\
     &   685.264 & 0.170 & 192.2382 & 0.0023 & 52450 &&  &&&&&\\
     &   685.065 & 0.044 & 192.1392 & 0.0007 & 52451 && {\bf GJ 9071} &&&&&\\
     &   683.084 & 0.038 & 192.0158 & 0.0006 & 52454 && 1--2 &  9971.76 & 0.20 & 239.5226 & 0.0009 & 52509 \\
     &   643.316 & 0.058 & 188.7846 & 0.0006 & 52509 &&      &  9972.00 & 0.26 & 239.5376 & 0.0012 & 52510 \\
     &   642.957 & 0.041 & 188.7072 & 0.0004 & 52510 &&      &  9924.03 & 0.24 & 240.0697 & 0.0011 & 52592 \\
\hline
 \end{tabular}
\end{table*}

\section{Astrometry}

The astrometric  measurements are presented in Table \ref{tab_res} where for each pair of
stars and epoch (MJD) the separation $\rho\,[mas]$ and the position angle 
$\theta$ [deg] is given. The $1 \sigma$ errors were calculated using the 
gaussian statistic of $\Delta x$ and $\Delta y$. The plate-scale uncertainty is
included into the separation error, but not into $\Delta \theta$ for Hale 
observations. It is because 
we wanted to show how small changes can be noticed between two nights where the same 
chip orientation is present (see GJ 860 1-2 in August -- MJD=52509 and 52510). 
CR orientation uncertainties are about one order of magnitude bigger, so they would 
dominate the $\theta$ error budget. The uncertainty in $\theta$ for GJ 300
is also underestimated because this system was observed with only one position 
of the field rotator of NIRC2. This is however not the case for 56~Per and GJ 569 B
where different values of the field rotator were used for dithering. Hence
the resulting formal error of the position angle presumably more
realistically reflects the accuracy of this AO system.
For systems observed more than once, the orbital, parallactic and proper motion can 
be seen. Even for the long-period binary GJ 195 \citep[$P \simeq$ 338 $yr$;][]{hein74} 
there is a clear signature of the orbital motion. Also a closer inspection of GJ 873 
reveals a motion of the double secondary system (see Fig. \ref{fig_rmsGJ}). 
GJ 873 B is probably a real binary but at a different distance from the Sun than 
GJ 873 A (parallactic motion is present). 

For 5 of our binaries (GJ 195, GJ 352, GJ 569 B, GJ~661 and GJ 860), 
the orbital solutions are known and can be found in the {\it Washington Double 
Star Catalog} (WDS)\footnote{{\it Sixth Catalog of Orbits of Visual Binary 
Stars} http://ad.usno.navy.mil/wds/orb6/orb6frames.html}. The corresponding 
orbital elements are presented in Table \ref{tab_elem}. The quality of the 
orbit is represented by the parameter $q$ -- the smaller value, the more 
accurate the orbital solution. In all five cases the orbital solutions are not 
perfect but only for GJ~195 the elements are poor due to a long period of
the binary.
 
In Figure \ref{fig_orb} the comparison between the orbit and our 
measurements is shown. There are several possible sources of the discrepancy 
between our data and the orbits: (1) the quality of the orbits; (2) the uncertainty 
in the pixel scale; (3) an imperfect ADR correction. The level of the
discrepancies ($\sim 100$ mas for GJ 195 and GJ 860, $\sim 20$ mas 
for GJ 352) seems to favor the first explanation. Also, if the cause was 
the ADR correction, one would expect a much higher scatter (see next section). 

\begin{figure*}
\includegraphics[width=0.87\textwidth]{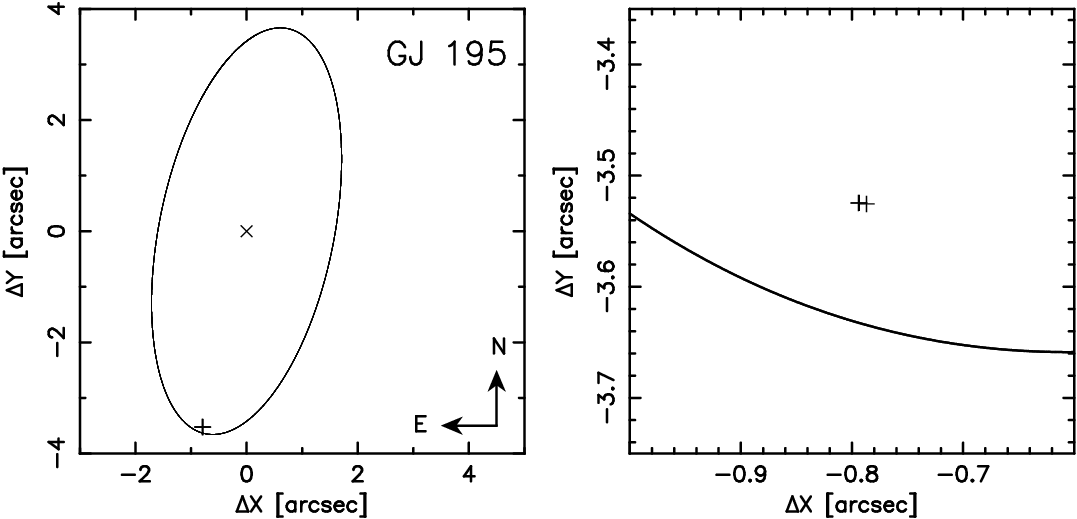}
\includegraphics[width=0.87\textwidth]{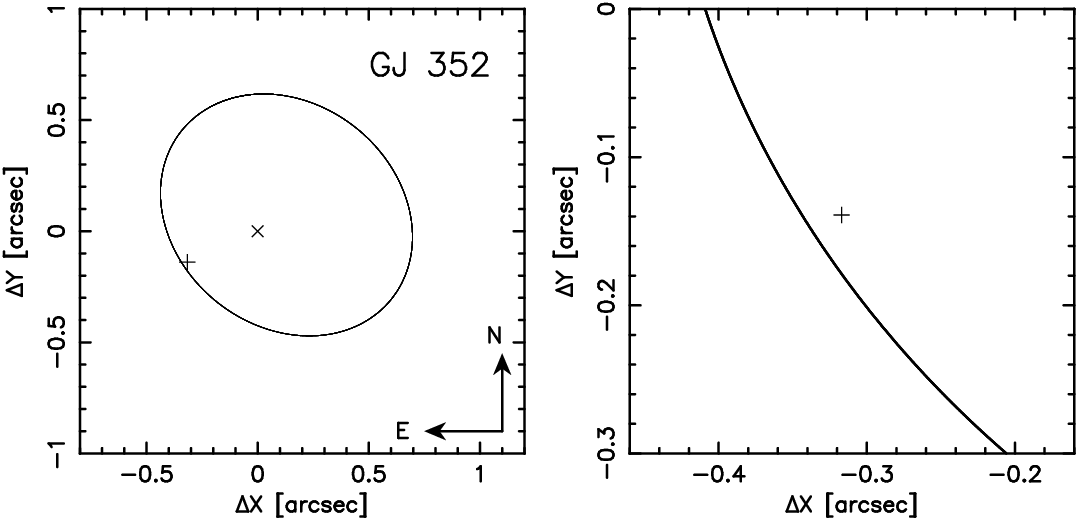}
\includegraphics[width=0.87\textwidth]{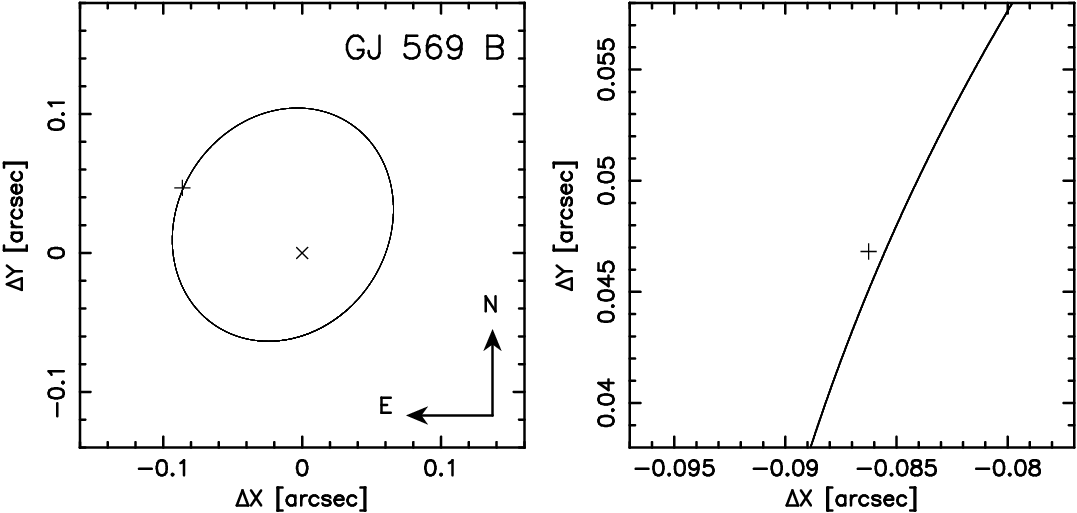}
\caption{Relative orbits from WDS compared with our astrometry. 
Right pannels are zoomed in on the secondary's position. The formal
error bars are smaller than the symbols used. The discrepancy is probably 
dominated by the quality of the orbital solutions.\label{fig_orb}}
\end{figure*}
 
\begin{figure*}
\includegraphics[width=0.87\textwidth]{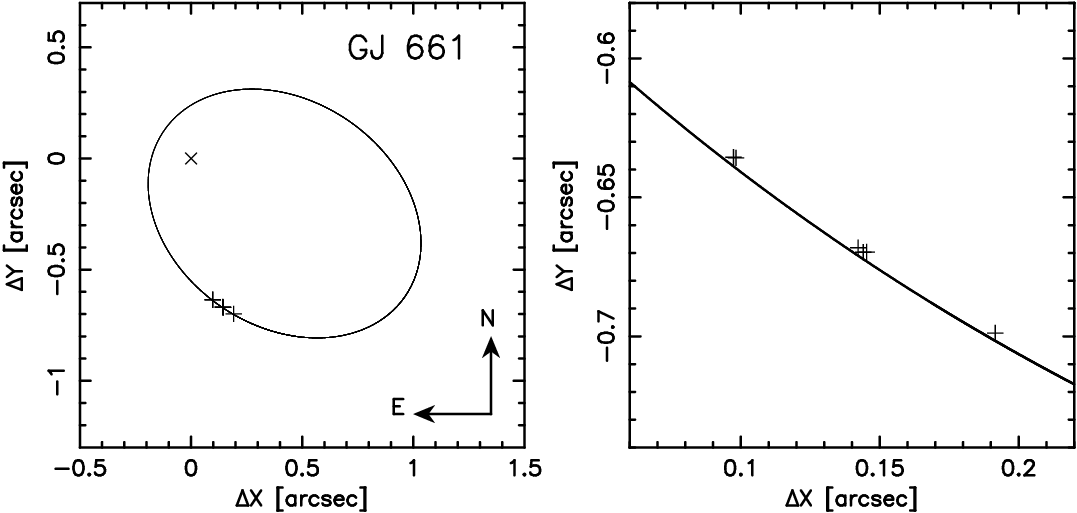}
\includegraphics[width=0.87\textwidth]{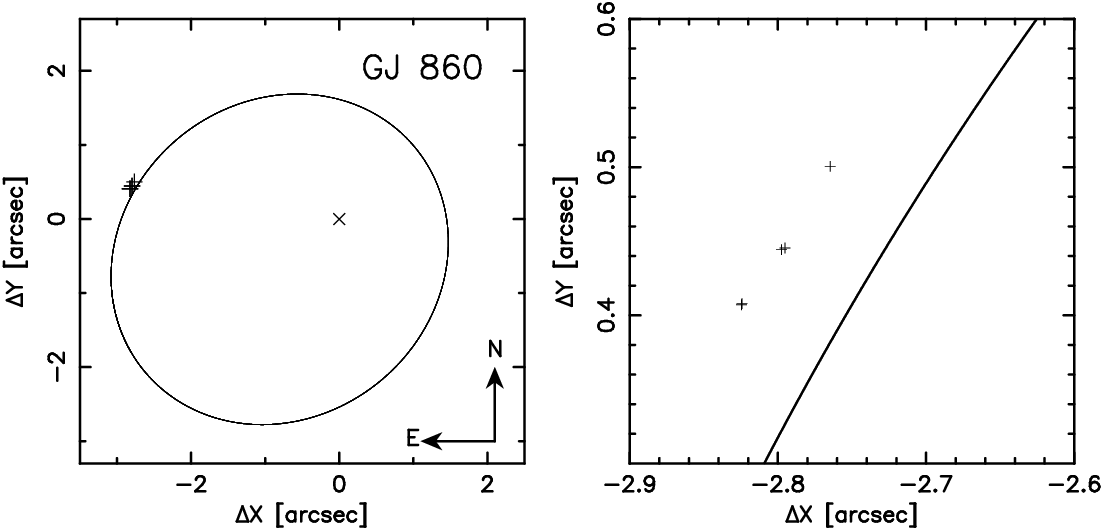}
\contcaption{}
\end{figure*}

\begin{table*}
 \caption{Orbital elements of five of our binaries for which orbital
solutions are available from WDS.\label{tab_elem}}
 \begin{tabular}{cccccccccc}
 \hline
 Star & P [yr] & a [mas] & e & i $[^\circ$] & $\Omega [^\circ]$ & $\omega [^\circ]$ & $\tau$ [MJD] & WDS ID & $q$ \\
 \hline \hline
GJ 195  &338.    &3720    &0.0     &65.     &168.5   &0.0	&55197   &05167+4600 &5\\
GJ 352  &18.4    &630     &0.29    &143.    &48.     &285.	&45663   &09313--1329&3\\
GJ 569B &2.424   &90.4    &0.312   &32.4    &321.3   &256.7	&51821   &14545+1606 &2\\
GJ 661  &12.9512 &762     &0.743   &149.14  &160.    &99.	&48373   &17121+4540 &2\\
GJ 860  &44.67   &2383    &0.41    &167.2   &154.5   &211.	&40666   &22280+5742 &2\\
 \hline
 \end{tabular}
\medskip 
\mbox{}
\end{table*}

\subsection{Overnight and long term astrometric precision}

During one night, for most of the binaries observed with the Hale telescope
we were able to go down below 500 $\mu$as in astrometric precision of 
$\rho$ and in some cases below 100. As expected, the 
precision is better for objects for which more single images were 
obtained. For pairs with similar brightness of the components, 
the astromertic error is smaller than for pairs with a high 
brightness difference. This is due to the poor S/N of the faint component
as well as a need not to saturate the bright one. It is imaginable that the 
usage of weighting might improve the precision a little.

In some cases when the stars are located on two different parts of the 
chip's mosaic, the astrometric errors are larger. In particular, 
for a very frequently observed, relatively close 
binary GJ 661 we achieved the highest overnight precision of 38 $\mu$as. 
Despite the fact that in this case the distortion correction is not 
perfect due to a location of the binary around the centre of the chip
where all 4 parts of the mosaic meet. \citet{met06} suggest not to use 
this area because of small differences in the chip's components geometry. 
It seems possible that for a binary like GJ 661, one may be able to 
achieve a precision even below our 38 $\mu$as in one night.

We achieved a similar level of overnight precision with the Keck II/NIRC2. The main 
difference between the two data sets is the significantly lower number of images. 
It is quite surprising that with 10 frames taken with the 'narrow' camera we 
reached $\sim 120$ $\mu$as precision for GJ 300. Nevertheless, one should treat 
this value cautiously. A low number of useful images does not allow for a
particularly accurate calibration.

Three systems GJ 661, GJ 860, GJ 873 and the open cluster NGC 6871 were observed more 
frequently than the remaining targets. Using their measurements, we can estimate 
the astrometric accuracy of the Hale telescope over a 120-140 day time span.
For the open cluster we take only pairs with star No. 5 which we believe is 
not a member of the cluster. Having 5 or 6 (for GJ 661) epochs, we can fit a 
2-nd order polynomial to the measured separations. Such a polynomial
is sufficient to model the proper, paralactic and orbital motion of
a close pair of stars. The fits are shown in Figure \ref{fig_rmsGJ}. 
The $rms$' for all fits are collected in Table \ref{tab_rms}. Note that again 
the resulting $rms$' are worse for pairs with very high brightness ratios such 
as those for GJ 860 with the star no. 3 and GJ 873 with the star no. 1. These $rms$' 
are also obviosuly higher than single night precisions for the corresponding
pairs of stars and can be treated as en estimate for the true astrometric
errors incorporating systematic effects due to an imperfect plate scale 
calibration, limited knowledge of the weather conditions and long term
astrometric stability of the telescopes/cameras which could not be
accounted for with our limited calibrations.
 
\begin{table}
\caption{The $rms$ of 2-nd order polynomial fits to the measurements of $\rho$ 
for the most frequently observed objects \label{tab_rms}}
 \begin{tabular}{rccc}
\hline
Pair & $rms$   & No. of & time span \\
     & [mas] & nights & [d]     \\
\hline \hline
GJ 661 1-2	& 0.282 & 6 & 122 \\
GJ 860 1-2	& 0.216 & 5 & 143 \\
       1-3	& 2.154 &&\\
       2-3	& 2.595 &&\\
GJ 873 1-2	& 1.127 & 5 & 143\\
       1-3	& 1.282 &&\\
       2-3	& 0.309 &&\\
NGC 6871 1-5	& 0.872 & 5 & 143\\
       2-5	& 0.717 &&\\
       3-5	& 0.763 &&\\
       4-5	& 2.671 &&\\
\hline
 \end{tabular}
\end{table}

\begin{figure*}
\includegraphics[width=0.86\textwidth]{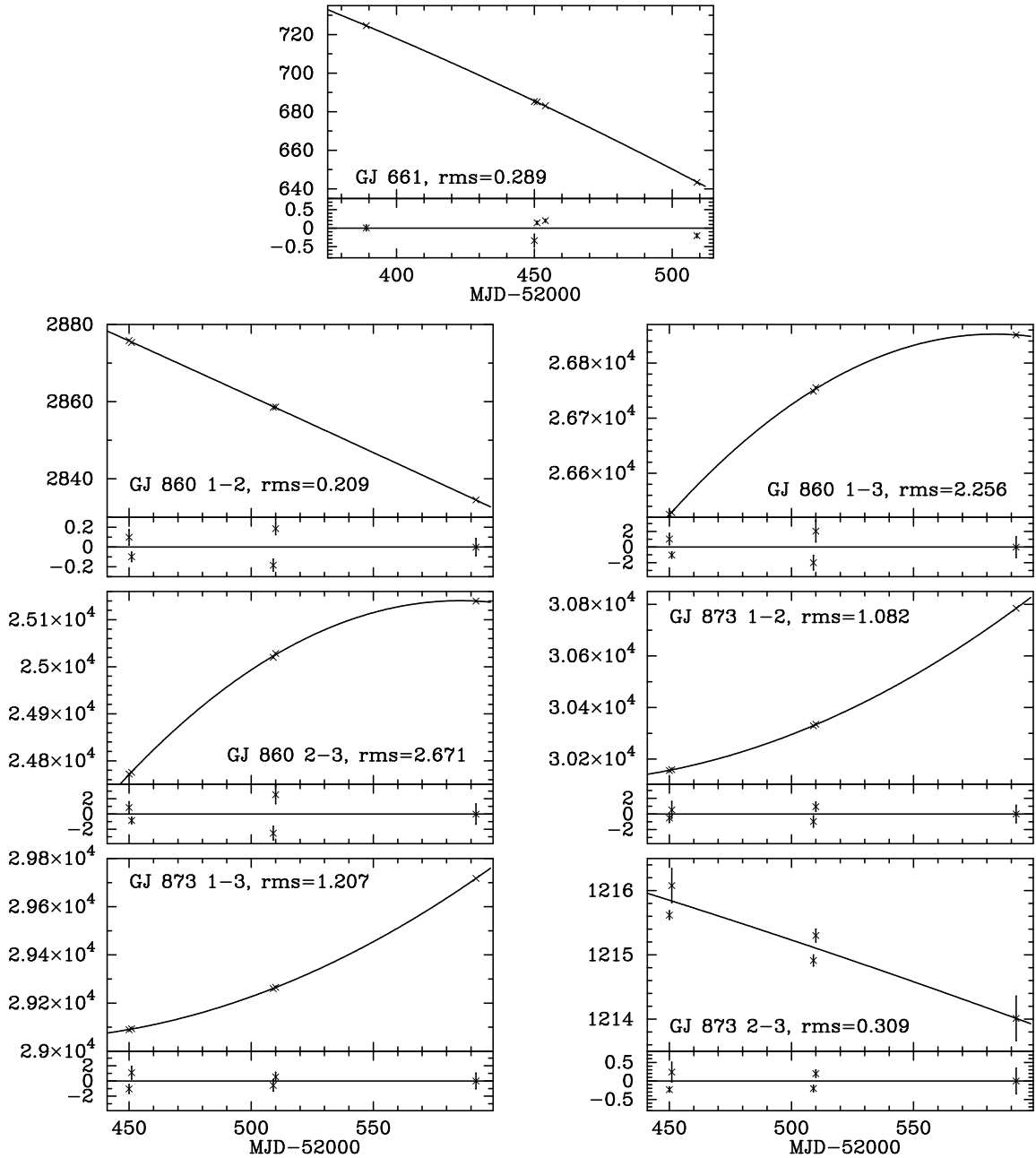}
\caption{A 2-nd order polynomial fits to the separations of stars 
 from Table \ref{tab_rms}. 
 \label{fig_rmsGJ}.}
\end{figure*}
\begin{figure*}
\includegraphics[width=0.86\textwidth]{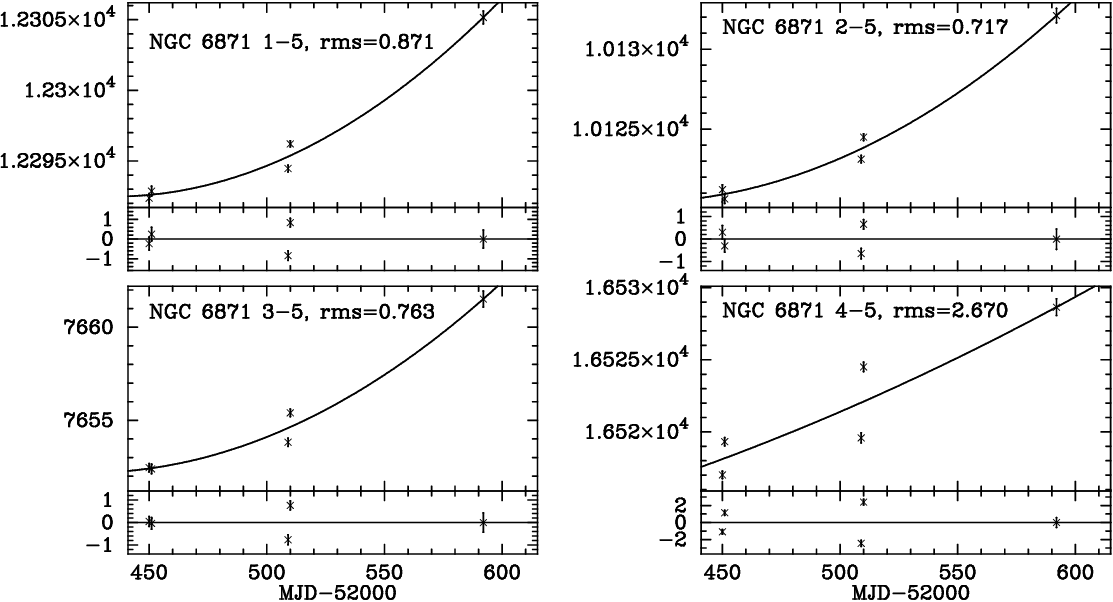}
\contcaption{}
\end{figure*}

\subsection{Detection limits}

The astrometric signal, $\Theta$ [$\mu$as], of a planet with a semi-major 
axis $a$ [AU] and mass $M_P$ (Jupiter masses) in a circular orbit around a star 
at a distance $d$ (parsecs) and mass $M_S$ (solar units) is given by \citep{pra96}
\begin{equation}\label{rel5}
\Theta = 1920 \frac{a}{d}\frac{M_P}{M_S}.
\end{equation}
Obviously the same relation can be used for an S-type planet\footnote{S-type, 
or satellite-type planet in binary/multiple is a planet 
orbiting only one of the components \citep{dvo84}.} 
in a wide binary system. Assuming that an astrometric signal above 
$3\sigma_{\rho}$ can be treated as a real one, $\Theta$ in the equation 
\ref{rel5}, may be replaced by $3\sigma_{\rho}$ [mas] (from Table \ref{tab_res}). 
After changing $d$ [pc] to parallax $\pi$ [mas] we obtain
\begin{equation}\label{rel6}
a M_P \,[AU \cdot M_J] = 1562.5 \frac{\sigma_{\rho}M_S}{\pi}.
\end{equation}
In the above $a M_P = 4$ [AU $\cdot$ M$_J$] means that we can detect 1 M$_J$ (or 
more massive) planet in 4 AU (or wider) orbit, or 2 M$_J$ planet in 2 AU orbit, 
etc. 

The detection limits for the binaries, in which at least one stellar 
mass is known or can be estimated are collected in Table \ref{tab_lim}. 
Subscripts $I$ and $II$ refer to the order of stars given in the first 
column. As one can see, in principle it is possible to achieve a sufficient 
astrometric precision to detect a massive planet or a brown dwarf with 
the Hale and Keck telescopes. As we have demonstrated in the
previous section, currently a longer term precision is up to several times 
lower than the one achieved over one night.
However, one can use the $rms$' from Table \ref{tab_rms} with the Equation \ref{rel6} 
and compute long-term planetary detection limits. These long-term
limits' are also listed in Table \ref{tab_lim}.

\section{Conclusions}

Nearby binary and multiple star systems are excellent targets for astrometric
searches for extrasolar planets thanks to their proximity and the availability 
of natural reference stars necessary for relative astrometry. In our study of 12 
visual binaries/multiples and 1 open cluster with the Hale and Keck II telescopes 
and their adpative optics facilities, we have demonstrated that over one night 
one is able to obtain an astrometric precision reaching $\sim$40
$\mu$as. Such a precison is sufficient to detect Jupiter mass planets 
around components of binary nad multiple stars. However, in order to turn
the precision into a long term accuracy required to detect planets, one must 
be able to account for ADR and the plate-scale changes. The ADR correction
requires accurate weather readings and the plate-scale changes must be
carefully calibrated. In our attempt to account for both, we were able
to achieve a long term (over a 140 days time span) accuracy ranging from
0.2 to 2.7 miliarseconds that is several times larger then the corresponding
overnight precision, but still allowing for detection of massive planets or 
brawn dwarfs. Since we have had limited means to carry out the
calibrations, it is quite possible that a higher long-term accuarcy can be
reached with the existing AO facilities.

\begin{table*}
\scriptsize
\caption{$a\,M_P$ limits.\label{tab_lim}}
 \begin{tabular}{ccccccccc}
 \hline
 Pair & $\pi$ & $M_I$ & $M_{II}$ & $\sigma_{\rho}$ & MJD & $a_IM_{P,I}$ & $a_{II}M_{P,II}$ & Ref.\\
      & [mas] & [M$_{\sun}$] & [M$_{\sun}$] & [mas] &    & [AU $\cdot$ M$_J$] & [AU $\cdot$ M$_J$] &   \\

 \hline
 \hline
GJ 195 1-2&     72.0&   0.53&   0.19&   0.31& 52509&   3.56&  1.28& 1\\ 
&               &       &       &       0.23& 52510&   2.65&  0.95&\\
&               &       &       &       0.12& 52592&   1.38&  0.50&\\

GJ 195 1-3&     &       0.53&    ---&   0.50& 52509&   8.28&  ---& 1\\
&               &       &       &       0.44& 52510&   5.06&  ---&\\
&               &       &       &       0.24& 52592&   2.76&  ---&\\

GJ 195 1-4&     &       0.53&    ---&   0.54& 52509&   6.21&  ---& 1\\
&               &       &       &       0.47& 52510&   5.41&  ---&\\
&               &       &       &       0.23& 52592&   2.64&  ---&\\

GJ 195 2-3&     &       0.19&    ---&   0.43& 52509&   1.77&  ---& 1\\
&               &       &       &       0.46& 52510&   1.90&  ---&\\
&               &       &       &       0.24& 52592&   0.99&  ---&\\

GJ 195 2-4&     &       0.19&    ---&   0.51& 52509&   2.10&  ---& 1\\
&               &       &       &       0.47& 52510&   1.94&  ---&\\
&               &       &       &       0.21& 52592&   0.87&  ---&\\

GJ 352 1-2&     94.95&  0.44&   0.41&   1.11& 52389&   8.04&  7.49& 2\\  

GJ 458 1-2&     65.29&  0.40&   0.37&   0.40& 52389&   3.83&  3.54& 3\\
&               &       &       &       0.28& 52450&   2.68&  2.48&\\
&               &       &       &       0.36& 52451&   3.45&  3.19&\\

GJ 507 1-2&     75.96&  0.46&   0.37&   0.45& 52389&   3.70&  3.43& 3\\
&               &       &       &       0.33& 52450&   3.12&  2.51&\\
&               &       &       &       0.36& 52451&   6.25&  5.03&\\

GJ 569B 1-2&    101.91& 0.071&  0.054&  0.11& 52337&  0.012&  0.009&4\\ 

GJ 661 1-2&     158.17&  0.379&  0.34&  0.079& 52389&   0.30&  0.29& 5\\   
&               &       &       &       0.170& 52450&   0.63&  0.62&\\
&               &       &       &       0.044& 52451&   0.17&  0.16&\\
&               &       &       &       0.038& 52454&   0.16&  0.15&\\
&               &       &       &       0.058& 52509&   0.22&  0.21&\\
&               &       &       &       0.041& 52510&   0.17&  0.16&\\
&               &       &       &       0.282& TOTAL&   1.05&  1.03&\\

GJ 767 1-2&     74.9&   0.44&   0.4&    0.17& 52509&   1.56&  1.42& 3\\
&               &       &       &       0.12& 52510&   1.10&  1.00&\\
&               &       &       &       0.09& 52592&   0.83&  0.75&\\

GJ 767 1-3&     &       0.44&   ---&    0.52& 52509&   4.77&  ---& 3\\
&               &       &       &       0.28& 52510&   2.57&  ---&\\
&               &       &       &       0.39& 52592&   3.58&  ---&\\

GJ 767 2-3&     &       0.4&    ---&    0.43& 52509&   3.59&  ---& 3\\
&               &       &       &       0.22& 52510&   1.83&  ---&\\
&               &       &       &       0.35& 52592&   2.92&  ---&\\

GJ 860 1-2&     249.53& 0.34& 0.2711&   0.078& 52450&   0.17&  0.13& 5\\   
&               &       &       &       0.048& 52451&   0.10&  0.09&\\
&               &       &       &       0.061& 52509&   0.13&  0.10&\\
&               &       &       &       0.061& 52510&   0.13&  0.10&\\
&               &       &       &       0.087& 52592&   0.18&  0.15&\\
&               &       &       &       0.216& TOTAL&   0.45&  0.37&\\

GJ 860 1-3&     &   0.34&   --- &       0.73& 52450&  1.55&   ---& 5\\
&               &       &       &       0.41& 52451&  0.86&   ---&\\
&               &       &       &       0.94& 52509&  2.01&   ---&\\
&               &       &       &       1.37& 52510&  2.95&   ---&\\
&               &       &       &       1.34& 52592&  2.86&   ---&\\
&               &       &       &       2.154& TOTAL& 4.60&   ---&\\

GJ 860 2-3&     & 0.2711&  ---  &       0.75& 52450&  1.28&   ---& 5\\
&               &       &       &       0.42& 52451&  0.71&   ---&\\
&               &       &       &       0.91& 52509&  1.29&   ---&\\
&               &       &       &       1.19& 52510&  2.02&   ---&\\
&               &       &       &       1.33& 52592&  2.54&   ---&\\
&               &       &       &       2.595& TOTAL& 4.96&   ---&\\

GJ 873 1-2&     &       0.36&   ---&    0.57& 52450&  1.62&   ---& 3\\
&               &       &       &       1.07& 52451&  3.04&   ---&\\
&               &       &       &       0.73& 52509&  2.07&   ---&\\
&               &       &       &       0.60& 52510&  1.71&   ---&\\
&               &       &       &       1.12& 52592&  3.20&   ---&\\
&               &       &       &       1.127& TOTAL& 3.29&   ---&\\

GJ 873 1-3&     198.07& 0.36&   ---&    0.58& 52450&  1.65&   ---& 3\\
&               &       &       &       0.85& 52451&  2.41&   ---&\\
&               &       &       &       0.76& 52509&  2.16&   ---&\\
&               &       &       &       0.61& 52510&  1.73&   ---&\\
&               &       &       &       1.03& 52592&  2.93&   ---&\\
&               &       &       &       1.282& TOTAL& 3.65&   ---&\\

GJ 9071 1-2&    72&     0.53&   0.49&   0.20& 52509&  2.22&  2.05& 3\\
&               &       &       &       0.26& 52510&  2.89&  2.67&\\
&               &       &       &       0.24& 52592&  2.66&  2.46&\\
 \hline
 \end{tabular}
\smallskip\\
\flushleft
References: (1) Fisher \& Marcy 1992; (2) S\"oderhjelm 1999; 
(3) Harmanec 1988; (4) Zapatero Osorio et. al 2004; (5) Delfosse et al. 2000.\\
Note: If $MJD$ is 'TOTAL', the limit refers to the $rms$ of the fit given in Table 
--- an estimate of a long-term astrometric precision for a given pair of
stars.
\ref{tab_rms}
\end{table*}

\section*{Acknowledgments}

M.K. is supported by the Foundation for Polish Science through a FOCUS
grant and fellowship. This work was supported by the Polish Ministry of 
Science and Higher Education through grants N203 005 32/0449 and 1P03D-021-29
and by NASA through grant NNG04GM62G.

This publication makes use of data products from the Two Micron All
Sky Survey, which is a joint project of the University of Massachusetts
and the Infrared Processing and Analysis centre/California Institute of
Technology, funded by the National Aeronautics and Space Administration
and the National Science Foundation.

\bsp

\label{lastpage}

\end{document}